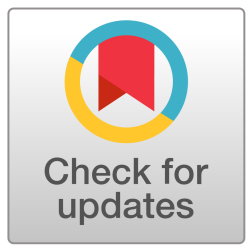

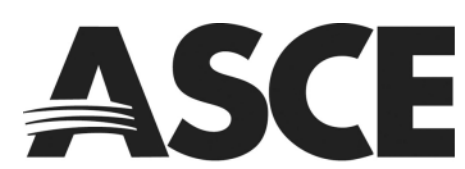

# Predicting the Fatigue Life of Asphalt Concrete Using Neural Networks

Jakub Houlík[1]; Jan Valentin[2]; and Václav Nežerka[3]

**Abstract:** The durability of asphalt concrete (AC) is determined in large part by its fatigue life. Traditional methods for assessing fatigue life are resource-intensive and time-consuming, and previous artificial neural networks (ANN) applications have not fully detailed their architectures nor utilized extensive data sets. Our study employs ANNs to predict AC fatigue life, focusing on strain level, binder content, and air-void content, and using a data set of 245 samples. We enhanced the ANN's predictive capability by optimizing hyperparameters and using mean square logarithmic error as the loss function. Our findings demonstrate improved fatigue life predictions, particularly highlighting the significant role of binder content. Additionally, we provide open access to our codes and data sets in a GitHub repository, promoting further research and model refinement. This approach advances the applicability and transparency of ANN models in predicting AC fatigue life.

## Introduction

Asphalt concrete (AC) is a composite material widely used in road construction. It consists of aggregates (crushed stone, gravel, or sand) and a bituminous binder, which is either used as paving-grade bitumen resulting from crude oil distillation or as a modified binder (e.g., polymer-modified bitumen) where various polymers or additives are mixed with the neat binder (Valentin et al. 2021).

The fatigue life of AC is a crucial parameter that determines to a large extent the maintenance costs (Isied and Souliman 2019). In this regard, AC should withstand cyclic heavy traffic loads under different environmental conditions and accommodate cumulative damage without excessive opening of cracks (Xiao et al. 2009; Shi et al. 2023). The main driver of AC deterioration and limited capability to withstand repeated tensile stresses is the aging of the bituminous binder due to oxidation (Bell et al. 1994; Sreedhar and Coleri 2020), and the failure occurs when the material cannot withstand the applied tensile strain (Beskou et al. 2016). While cold temperatures lead to increased brittleness and, in case of freezing, ice causes the opening of cracks (Maczka and Mackiewicz 2023), elevated temperatures make bituminous binders more viscous and prone to flow (Moreno-Navarro et al. 2017).

Fatigue resistance is influenced by several factors, most notably the binder content (Maczka and Mackiewicz 2023), air voids content (Enríquez-León et al. 2021; Praticò and Moro 2012), loading strain (Liao et al. 2012; Zhang et al. 2023), and operational temperature (Yang et al. 2023; Zhang et al. 2023). It is typically evaluated using different laboratory methods, most commonly four-point bending (4PB), semicircular bending, and indirect tensile tests. The 4PB fatigue test assesses fatigue behavior under bending loads, determining fatigue life, stiffness, and strength (Cheng et al. 2021, 2022a; Sudarsanan and Kim 2022). The semicircular test, similar in nature, allows, in addition, evaluation of fracture toughness (Saha and Biligiri 2015; Zielinski 2023). The indirect tensile test applies repeated tensile loads to cylindrical AC specimens using a diametral compression setup, measuring the number of cycles to failure ($N_f$) or accumulated strain at a constant load frequency and amplitude (Nguyen et al. 2012; Sudarsanan and Kim 2022; Vishnu 2023).

Understanding the relationships between fatigue life and AC composition, climatic conditions, and loading rate is vital for a proper mix design, but extensive testing is time-consuming. The standardized testing procedures require 14–42 days for the preparation of specimens and proper curing (CEN 2018), and the test itself takes from hours to several days, depending on the load level and frequency of loading. While the standardized testing methods offer valuable insights, they are time- and resource-intensive. Leveraging machine learning (ML) models and already collected data, AC fatigue life can be predicted when time-consuming testing is impossible.

Numerous studies have extensively reviewed various aspects such as the underlying mechanisms, characterization techniques, and mitigation strategies, highlighting the intricate interactions between AC's components and the influence of external factors such as aging and environmental conditions on fatigue behavior (Xiao et al. 2009; Isied and Souliman 2019; Wang et al. 2023; Lundstrom et al. 2004). Recent studies on ML-assisted mix design (Chen and Liu 2022; Hooda et al. 2022; Lian et al. 2022) have laid the groundwork for developing models that integrate a broader range of variables to predict fatigue life. Although some critical factors, such as asphalt type and aggregate type, have been noted as significant, our model focuses on binder content (%) and air voids (%), which are consistently reported across studies and critical for the longevity of AC pavements, as identified by Alnaqbi et al. (2024). The exclusion of asphalt and aggregate types was a deliberate decision due to the vast variability in these materials, which would require an extensive testing agenda beyond the scope of this study. Nevertheless, by making our data sets and software open source, we provide a foundation for future research to refine and expand the model to include these parameters if desired. To the best of our knowledge, such a comprehensive application of neural networks for predicting fatigue life in laboratory specimens, complete with open access to developed tools, has not previously been reported in the scientific literature.

[1]Ph.D. Student, Dept. of Physics, Faculty of Civil Engineering, Czech Technical Univ. in Prague, Thákurova 7, Praha 6 166 29, Czech Republic. ORCID: https://orcid.org/0009-0001-5956-026X. Email: jakub.houlik@fsv.cvut.cz

[2]Associate Professor, Dept. of Road Structures, Faculty of Civil Engineering, Czech Technical Univ. in Prague, Thákurova 7, Praha 6 166 29, Czech Republic. ORCID: https://orcid.org/0000-0001-9155-1921. Email: jan.valentin@fsv.cvut.cz

[3]Associate Professor, Dept. of Physics, Faculty of Civil Engineering, Czech Technical Univ. in Prague, Thákurova 7, Praha 6 166 29, Czech Republic (corresponding author). ORCID: https://orcid.org/0000-0001-7360-0507. Email: vaclav.nezerka@cvut.cz

## Methodology

Our study's methodology involved using artificial neural networks (ANNs) to predict fatigue cycle quantification in AC specimens. We based our predictions on existing data sets obtained exclusively during 4PB tests loaded at a constant strain rate. This approach was chosen because it is widely adopted in the field, thereby ensuring the availability of an extensive and robust data set for model training and validation.

### *Data Collection*

The data employed for training and testing were derived from four selected peer-reviewed scientific publications, each chosen for its consistent testing methodology (4PB 20°C, 10 Hz), thereby ensuring comparability despite the limited number of sources. These publications provided essential parameters such as binder content (%), air voids (%), loading strain ($\varepsilon$), test temperature, loading frequency ($f$), and the number of cycles until failure ($N_f$). Our data set exclusively comprises samples subjected to 4PB at temperatures of 20°C and 70°F (21.1°C). In total, the analysis incorporated 245 samples from four distinct studies (Table 1). To present a comprehensive overview of the data extracted from the literature, Table 1 also includes entries that do not align with our selection criteria, marked as "No." Although these samples were excluded from the final modeling data set, retaining them in the table underscores the full breadth of our initial data collection and facilitates the possibility of future meta-analyses or alternative modeling approaches where these samples could be relevant.

The samples in the selected studies were prepared and tested under laboratory conditions but differed slightly, offering variability to the data set. Li et al. (2012) utilized aggregates with a maximum nominal size of 19 mm and tested two types of binder: PG 64-22 and PG 76-22. Saboo et al. (2016) evaluated three types of mixes: bituminous concrete, dense bituminous macadam, and aggregates bonded with mastic asphalt. They incorporated four types of binders, including modifications with styrene-butadiene-styrene (SBS), and ethylene acetate, targeting an air void content of 4%. Pais et al. (2002) also used granitic aggregates with a maximum nominal size of 19 mm and explored three types of bituminous binders: 10/20, 35/50, and SBS 50/70. Similarly, Witczak et al. (2013) employed aggregates with the same maximum nominal size of 19 mm and tested three types of bituminous binders: PG 58-28, PG 64-22, and PG 76-16.

The original distribution of $N_f$ exhibited significant skewness, primarily due to the long tail of the data (Figs. 1 and 2). To address this and ensure the data set was suitable for outlier detection, we applied a Box–Cox transformation, which normalizes data by finding the optimal transformation parameter $\lambda$ (Box and Cox 1964; Hinkle and Emptage 1991; Sakia 1992), as follows:

$$N_f^{BC}(\lambda) = \frac{N_f^{\lambda} - 1}{\lambda} \quad (1)$$

where $\lambda$ = transformation coefficient. The optimal $\lambda$ value for our data set was identified as $\lambda = -0.09$.

To assess the normalization effect, we performed the Shapiro–Wilk test (Shapiro and Wilk 1965) and analyzed Q-Q plots, comparing the original data, log-transformed data, and Box–Cox-transformed data (Fig. 1). The original data exhibited significant skewness and heavy tails, as reflected in the Q-Q plot and the long-tailed histogram. The logarithmic transformation reduced skewness but did not fully achieve normality, as indicated by the persistent deviations in the Q-Q plot and low $p$_value. The Box–Cox transformation, with $\lambda = -0.09$, yielded the most normalized distribution, closely aligning with the theoretical normal line in the Q-Q plot and producing a histogram with minimal skew.

After transforming the data, we employed a $Z$-score to detect potential outliers, as follows:

$$Z_i = \frac{N_{f,i} - \mu}{\sigma} \quad (2)$$

where $N_{f,i}$ = individual data point; $\mu$ = the mean; and $\sigma$ = standard deviation of the transformed data set. The $Z$-score threshold was selected as 3.0 based on recommendations by Chandola et al. (2009) and Seo (2006), ensuring alignment with commonly accepted practices. While this threshold is common across various disciplines, its use in fatigue-related studies is also supported by field-specific literature. For instance, Shaygan and Jamshidi (2023) applied the value of 3.0 ($\pm 3\sigma$) for $Z$-score filtering to eliminate extreme geomechanical data with logarithmic distribution.

**Table 1.** Summary of studies on AC fatigue measurements detailing data utilization in our modeling and reasons for exclusion of specific data sets

| Reference | Binder content (%) | Air voids (%) | $\varepsilon$ ($\times 10^{-6}$) | Temperature (°C) | $f$ (Hz) | $N_f$ ($\times 10^3$) | Samples | Used for modeling |
|---|---|---|---|---|---|---|---|---|
| Li et al. (2012) | 4.0–6.0 | 1.3–7.7 | 500–1,000 | 20 | 10 | 4–1,588 | 22 | Yes |
| Pais et al. (2002) | 4.7–6.5 | 1.2–12.8 | 115–800 | 20 | 10 | 3–19,500 | 115 | Yes |
| Saboo et al. (2016) | 4.7–6.7 | 4.0–4.1 | 400–1,000 | 20 | 10 | 1–152 | 38 | Yes |
| Witczak et al. (2013) | 4.2–5.2 | 3.9–11.3 | 100–500 | 4.4, 21.1, 37.8 | 10 | 2–700 | 229 | Yes[a] |
| Ameri et al. (2016) | N/A | N/A | 400–1,000 | 20 | 10 | 5–490 | 32 | No[b] |
| Bańkowski (2018) | 3.7–6.3 | 2.1–4.3 | N/A | 10 | 10 | N/A | 5 | No[c] |
| Al-Khateeb and Ghuzlan (2014) | 5.1 | 4.0 | N/A | 20, 30 | 3, 5, 8, 10 | N/A | 40 | No[c] |
| Harvey et al. (1995) | 4.0–6.0 | 1.2–8.9 | 150–300 | 19 | 10 | 14–1,472 | 30 | No[c] |
| Lv et al. (2023) | N/A | 3.4–4.7 | N/A | 20 | 10 | 0–3,365 | 72 | No[c] |
| Maczka and Mackiewicz (2023) | 3.8–4.2 | 4.0–6.2 | N/A | N/A | N/A | 149–1,958 | 12 | No[b] |
| Pakholak et al. (2021) | 6.3–12.0 | 2.9–23.8 | N/A | 5, 15, 25, 35 | 10 | N/A | 96 | No[c] |
| Shen and Lu (2011) | N/A | N/A | 300–1,000 | 20 | 10 | 3–446 | 5 | No[b] |
| Shu et al. (2008) | 5.0–5.5 | 7.0 | 600 | 25 | 10 | 83–117 | 4 | No [d] |
| Yan et al. (2018) | 4.0–7.0 | 2.2–6.2 | 500–1,500 | 5 | N/A | 0–37 | 54 | No[d] |
| Zeiada et al. (2013) | 4.2–5.2 | 3.4–10.0 | 75–200 | 20 | 10 | 3–676 | 21 | No[c] |
| Zhang et al. (2013) | 6.5–6.6 | 6.0 | 202–1,032 | 7, 13, 20 | 10 | 1–230 | 38 | No[c] |
| Zhou et al. (2013) | 5.8–7.3 | 3.5–4.5 | 75–500 | 25 | 10 | 0–43,900 | 24 | No[c] |
| Whole data set for modeling | 4–6.7 | 1.2–12.8 | 115–1,000 | 20 | 10 | 0–19,500 | 245[a] | — |

[a]Only specimens tested at 70°F from Witczak et al. (2013) were considered, ensuring consistency with our selected temperature/frequency parameters.
[b]Missing information about binder content and air voids or temperature.
[c]Different testing method than 4PB.
[d]Different testing temperature than 20°C/70°F.

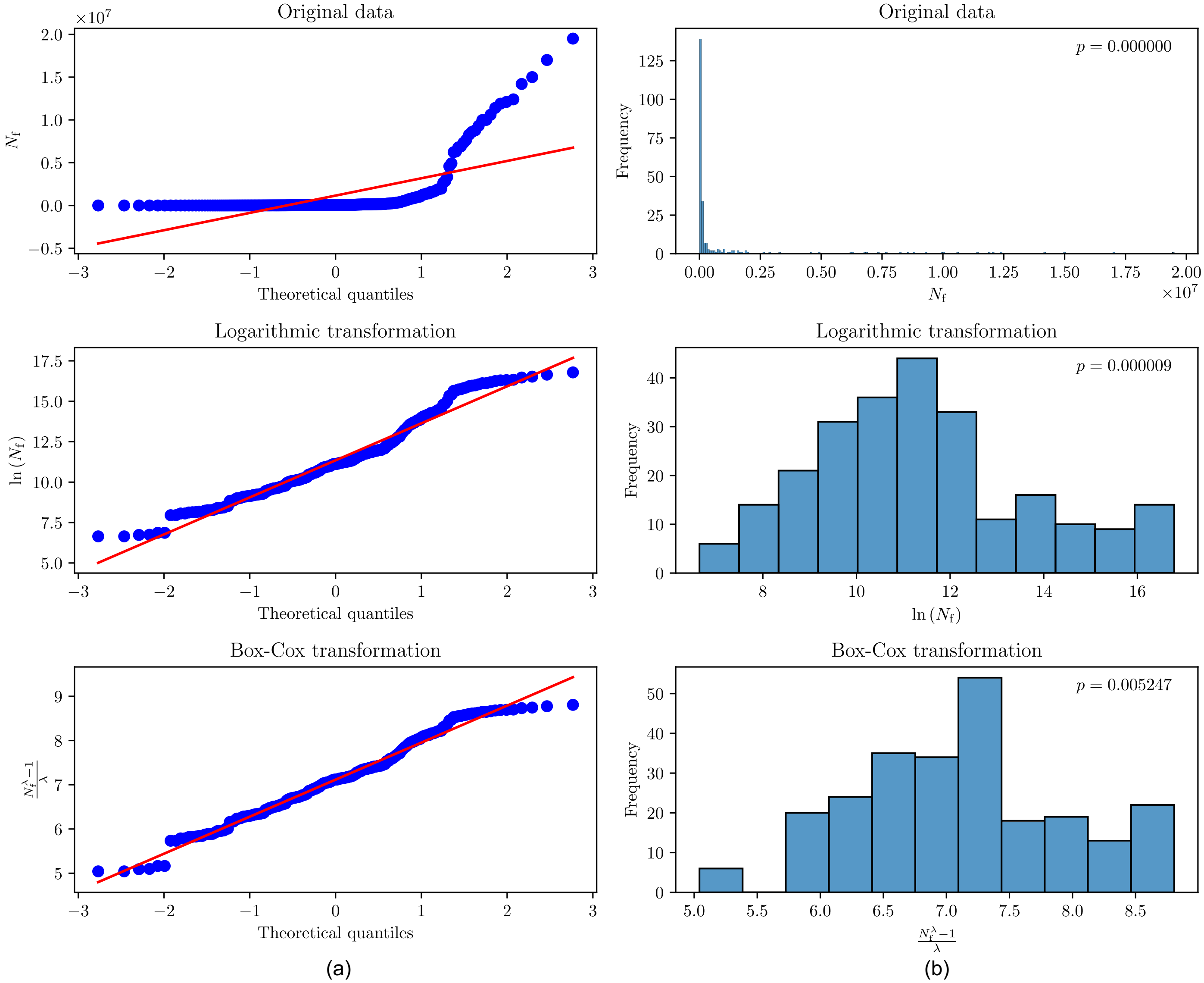

**Fig. 1.** Q-Q plots: (a) and distribution histograms; and (b) for the $N_f$ data set before and after transformations, comparing the data set to a theoretical normal distribution.

Notably, no data points exceeded the $Z$-score threshold, indicating that no outliers were present in the transformed data set. As a result, the original data set was retained without additional filtering, ensuring all available data were utilized for model training and evaluation.

To enhance the robustness of outlier detection, we complemented our approach, based on standard $Z$-scores thresholding after Box–Cox transformation, with the modified $Z$-score method (Dastjerdy et al. 2023). The modified $Z$-score is a robust method for outlier detection that uses the median and the median absolute deviation (MAD), making it more resilient to skewed distributions and extreme values than the standard $Z$-score. For a given data point $N_{f,i}$, the modified $Z$-score is calculated as

$$Z_{\mathrm{mod},i} = \frac{0.6745(N_{f,i} - \widetilde{N_f})}{\mathrm{MAD}} \tag{3}$$

where $\widetilde{N_f}$ = median of the transformed data set; and MAD = median($|N_{f,i} - \widetilde{N_f}|$) is the median absolute deviation. Even after applying a threshold $Z_{\mathrm{mod},i} > 3.5$, as recommended by Iglewicz and Hoaglin (1993), no data points were identified as outliers. This result aligns with the findings from the conventional $Z$-score method applied to the Box–Cox transformed data, confirming that all input data could be retained for modeling without the need for exclusion.

The final distributions and statistics of the collected data are summarized in Table 2, while pair plots with distributions for individual features are provided in Fig. 2. The distribution of $N_f$ (Fig. 2) reveals a pronounced skew, with the majority of values concentrated below $0.2 \times 10^6$, accompanied by a long tail extending toward higher fatigue life ranges. It is also clear that the features are independent of each other (low $\rho$) and that $N_f$ decreases with the increments of $\varepsilon$. Table 2 provides detailed descriptive statistics for binder content, air voids, $\varepsilon$, and the $N_f$, including the minimum, maximum, first quartile (Q1), third quartile (Q3), mean, standard deviation and median, ensuring a comprehensive understanding of the data set.

### *ANN*

The ANN model is highly effective for processing tabular numerical data, which makes it well-suited for fatigue life predictions.

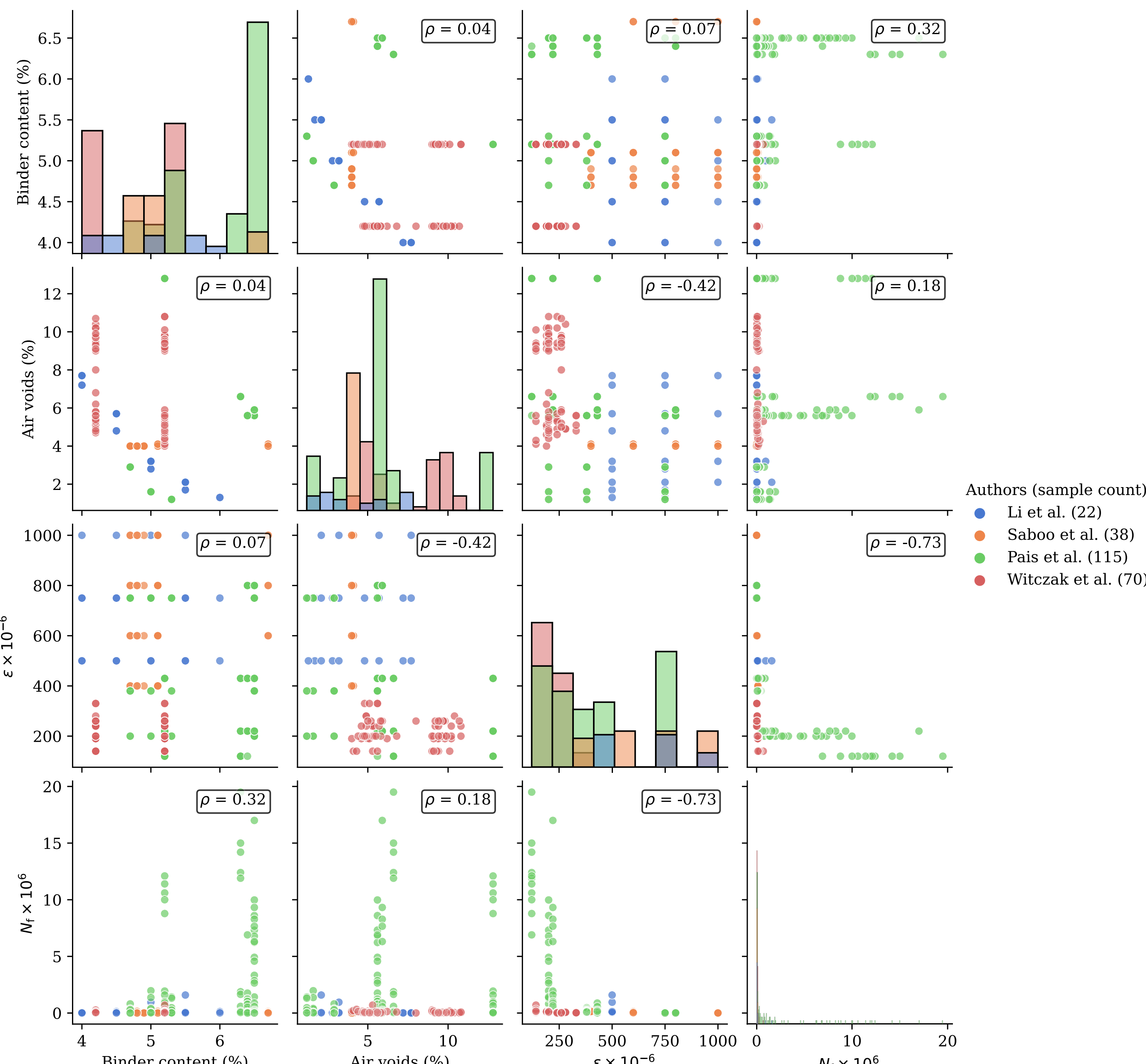

**Fig. 2.** Pairwise scatter plots and distributions of collected data, with features split by individual studies. The $N_f$ distribution exhibits a significant concentration of values below $0.2 \times 10^6$, along with a long tail representing higher fatigue life values; Spearman's correlation coefficients ($\rho$) are shown for all pairwise relationships, highlighting monotonic associations between features.

The model assigns weights to different input variables and combines them through several hidden layers to generate the output (Fig. 3). These weights are adjusted based on validation data through multiple iterations to improve the model's accuracy.

**Table 2.** Descriptive statistics of the collected data set, including binder content (%), air voids (%), $\varepsilon$ in $\times 10^{-6}$, and $N_f$ in $\times 10^3$; the statistics provide insights into the range, central tendency, and variability of the data set

| | Binder content (%) | Air voids (%) | $\varepsilon$ ($\times 10^{-6}$) | $N_f$ ($\times 10^3$) |
|---|---|---|---|---|
| Minimum | 4.0 | 1.2 | 115 | 0.8 |
| Q1 | 4.7 | 4.0 | 200 | 15.4 |
| Q3 | 6.4 | 6.6 | 750 | 255 |
| Maximum | 6.7 | 12.8 | 1,000 | 19,500 |
| $\mu$ | 5.4 | 5.9 | 426 | 1,136 |
| $\sigma$ | 0.84 | 2.85 | 263 | 3,075 |
| Median | 5.2 | 5.6 | 375 | 67.6 |

We developed the prediction model using the sequential model from the Keras library, version 2.12.0rc1. Before training, the data were normalized using the *MinMaxScaler* function from the *sklearn.preprocessing* module, scaling the input variables between 0 and 1. The sequential model was configured with the following hyperparameters: number of hidden layers ($n_h$), number of neurons in each hidden layer ($n_{neur,h}$), the activation function, and the optimization algorithm (Table 3). We systematically varied these hyperparameters by minimizing the selected loss function during training, with $R^2$ serving as the primary metric for evaluating each model's performance. The coefficient of determination, $R^2$, is defined as

$$R^2 = 1 - \frac{\sum_{i=1}^{n} (N_{f,i} - \hat{N}_{f,i})^2}{\sum_{i=1}^{n} (N_{f,i} - \bar{N}_f)^2} \tag{4}$$

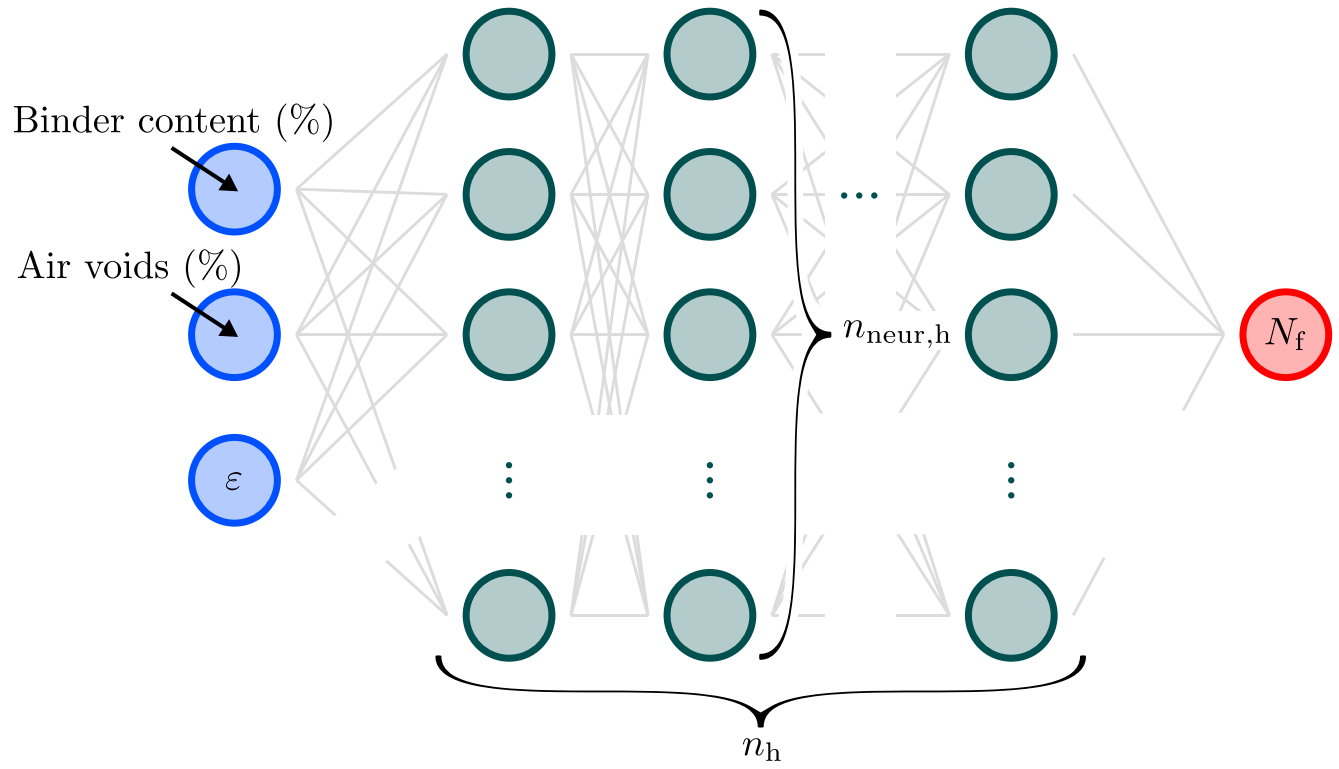

**Fig. 3.** Scheme of the ANN architecture used for modeling, illustrating the network parameters.

**Table 3.** Hyperparameters tested during model development and values selected for final model

| Hyperparameter | Grid | Selected |
|---|---|---|
| Loss function | $L_{\mathrm{MSE}}$, $L_{\mathrm{LMSE}}$ | Both |
| Optimizer | Adam, Nadam, RMSprop | RMSprop |
| Activation function | ReLU, linear, sigmoid | ReLU |
| $n_{\mathrm{h}}$ | 1, 2, 3, 4 | 2 |
| $n_{\mathrm{neur,h}}$ | 10, 50, 100, 150, 200 | 200 |

where $N_{\mathrm{f},i}$ = measured output; $\hat{N}_{\mathrm{f},i}$ = model prediction; $\bar{N}_{\mathrm{f}}$ = mean of $N_{\mathrm{f},i}$; and $n$ = number of samples.

In addition to $R^2$, we further quantified performance using the adjusted $R^2$ ($R^2_{\mathrm{adj}}$) and the scatter index (SI). The $R^2_{\mathrm{adj}}$ incorporates the number of predictors $p$ in the model, as follows:

$$R^2_{\mathrm{adj}} = 1 - \frac{(1 - R^2)(n - 1)}{n - p - 1} \tag{5}$$

This helps assess how well the model generalizes without overfitting. SI illustrates how model errors compare to the overall scale of the data set, as follows:

$$\mathrm{SI} = \frac{\sqrt{\frac{1}{n}\sum_{i=1}^{n}(N_{\mathrm{f},i} - \hat{N}_{\mathrm{f},i})^2}}{\frac{1}{n}\sum_{i=1}^{n} N_{\mathrm{f},i}} \tag{6}$$

where the numerator is the root mean square error and the denominator is the mean of the measured data.

Throughout the training, we focused on minimizing the loss function, ultimately saving the version of the model that achieved the lowest loss for validation data, representing the most accurate predictions without overfitting. Given the wide range of $N_{\mathrm{f}}$ data, which are typically represented on a logarithmic scale to accommodate their vast spread (Bańkowski 2018; Moreno-Navarro et al. 2017), we employed two distinct loss functions. The first is the mean square error ($L_{\mathrm{MSE}}$), defined as

$$L_{\mathrm{MSE}}(N_{\mathrm{f}}, \hat{N}_{\mathrm{f}}) = \frac{1}{n}\sum_{i=1}^{n}(N_{\mathrm{f},i} - \hat{N}_{\mathrm{f},i})^2 \tag{7}$$

and the mean square logarithmic error ($L_{\mathrm{MSLE}}$), defined as

$$L_{\mathrm{MSLE}}(N_{\mathrm{f}}, \hat{N}_{\mathrm{f}}) = \frac{1}{n}\sum_{i=1}^{n}(\log(N_{\mathrm{f},i} + 1) - \log(\hat{N}_{\mathrm{f},i} + 1))^2 \tag{8}$$

This strategic choice aimed to optimize our models for the logarithmic nature of fatigue life variability. Employing both $L_{\mathrm{MSE}}$ and $L_{\mathrm{MSLE}}$ as loss functions enabled us to fine-tune the model's sensitivity to the scale of the data.

The ANN was trained for 200 000 (when using $L_{\mathrm{MSE}}$) or 100 000 (when using $L_{\mathrm{MSLE}}$) epochs to ensure a sufficient number of iterations for the learning process.

To ensure the robustness and reliability of our ANN model, we split the data set into a test set (25%) and a training set (75%). The 75% training portion was further divided such that 20% of it was used for validation and 80% for training. In addition, we employed a four-fold cross-validation approach on the training set, using the *sklearn.model_selection.KFold* class. In each cross-validation cycle, one fold was used for validation, while the remaining three folds were used for training. This process was repeated for all folds, minimizing the risk of overfitting to particular data subsets and ensuring a thorough evaluation. We opted for four-fold cross-validation to strike a balance between computational efficiency and maintaining sufficient data in each fold. Moreover, the data set was shuffled prior to splitting with the *KFold* function, guaranteeing a randomized and uniform distribution of data across folds.

The modeling and subsequent training were carried out on a Dell XPS 9870 notebook with the following specifications: Intel (R) Core(TM) i7-8550U CPU (1.80 GHz), 16 GB RAM, PM981 NVMe Samsung SSD hard drive, and Windows 11 operating system. The code was developed in Python, version 3.10.14. With such hardware, training a single model for 200 000 epochs took between 1 and 1.5 h.

## Results and Discussion

### *Hyperparameters*

We tested multiple configurations before fine-tuning the ANN hyperparameters and observed that ReLU consistently yielded better performance. While rectified linear unit (ReLU) reduces the vanishing gradient problem, it can suffer from "dying ReLU," where neurons become inactive due to large negative weights or biases. We coupled ReLU with the root mean square propagation (RMSprop) optimizer. Guided by these insights, we conducted a detailed examination to determine the optimal settings for $n_{\mathrm{h}}$ and $n_{\mathrm{neur,h}}$, testing these configurations against both $L_{\mathrm{MSE}}$ and $L_{\mathrm{MSLE}}$ loss functions (Figs. 4 and 5, respectively). This analysis revealed that a structure with $n_{\mathrm{h}} = 2$ and $n_{\mathrm{neur,h}} = 200$ achieved the highest $\overline{R^2}$ score after averaging the $R^2$ score of all four folds.

Subsequent evaluations of various activation functions under this configuration affirmed that ReLU was the superior choice. Linear activation functions led to compromised predictions, while sigmoid resulted in $\overline{R^2}$ around 0, indicating inadequate model performance. This outcome underscores ReLU's advantage in maintaining nonlinearity in deep networks, which is crucial for tackling fatigue life prediction.

When comparing the efficacy of different optimizers, while the impact was less pronounced, the RMSprop optimizer consistently showed superior performance when using the ReLU activation function for both $L_{\mathrm{MSE}}$ and $L_{\mathrm{MSLE}}$ (Fig. 6). RMSprop adjusts the learning rate by using a moving average of the squared gradients, providing stability in noisy data environments by smoothing out the updates (Peng and Lee 2024). The Adam optimizer leverages adaptive learning rate to enhance weight adjustments (Jentzen and Riekert 2023; Kingma and Ba 2015). On the other hand, Nadam, which integrates Nesterov momentum into the Adam optimizer's structure, speeds up the model updates (Cheng et al. 2022b). This approach can sometimes overshoot the smallest or optimal values or struggle to stabilize on complex, uneven loss surfaces. While sometimes yielding faster convergence, this aggressive update strategy can lead to instability and lower $R^2$ scores, mainly when dealing with nonuniform data distributions or outliers.

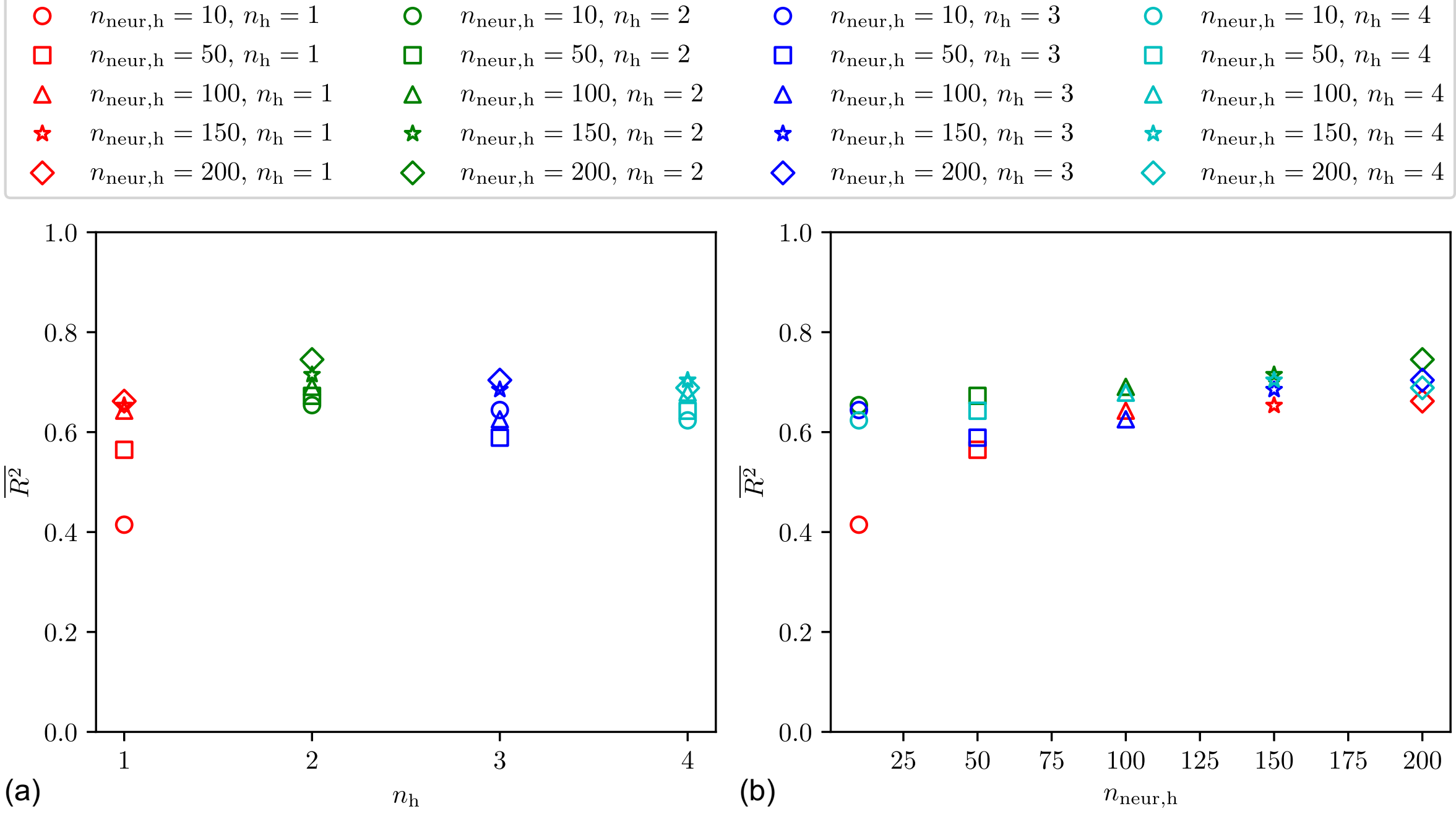

**Fig. 4.** Variation of the model performance, measured by the $\overline{R^2}$ score (mean for different folds), in relation to: (a) $n_{\mathrm{h}}$ and (b) $n_{\mathrm{neur,h}}$ (right) when employing $L_{\mathrm{MSE}}$ as the loss function; RMSprop optimizer and ReLU activation function.

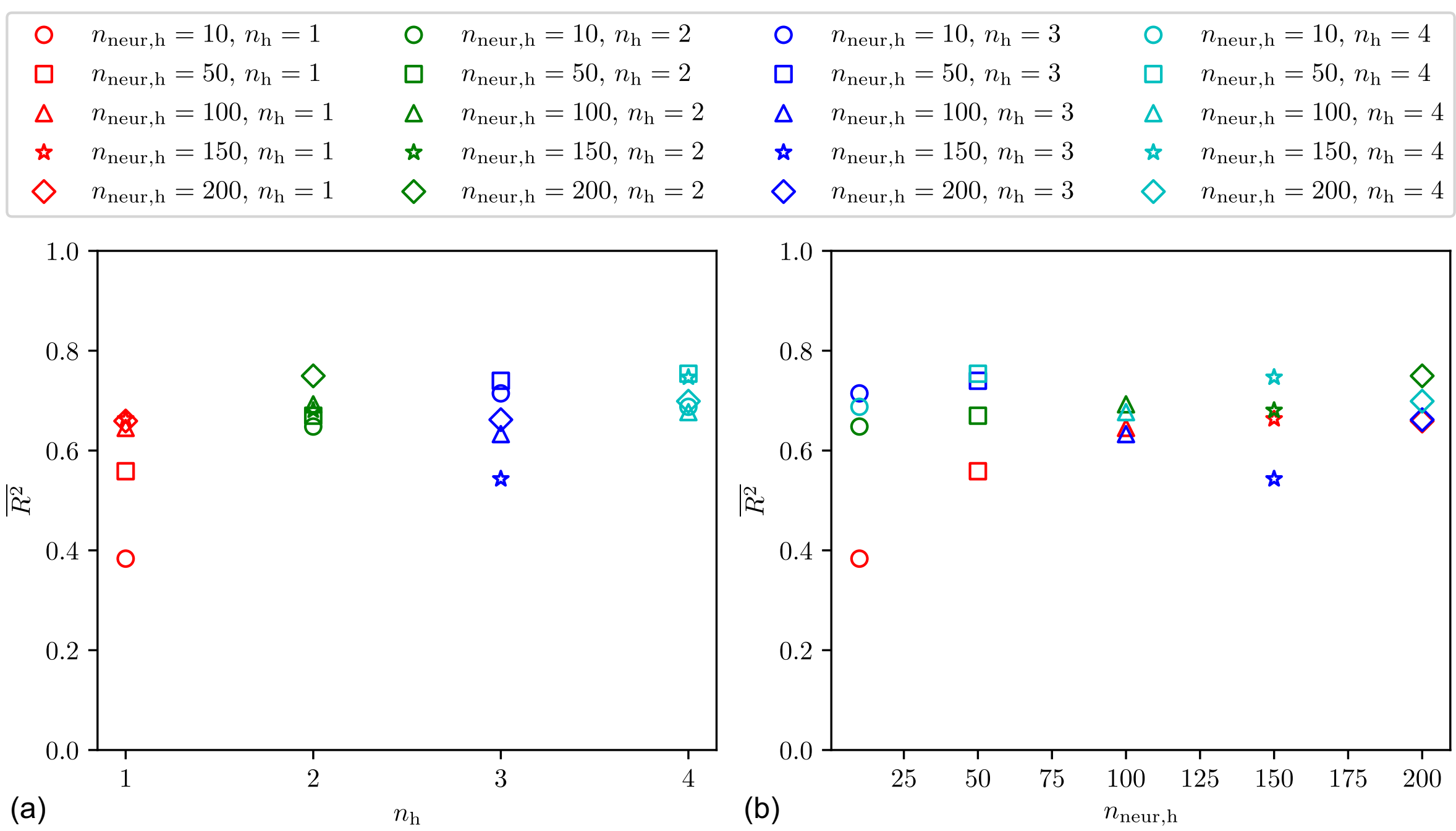

**Fig. 5.** Variation of the model performance, measured by the $\overline{R^2}$ score (mean for different folds), in relation to: (a) $n_{\mathrm{h}}$ and (b) $n_{\mathrm{neur,h}}$ when employing $L_{\mathrm{MSLE}}$ as the loss function; RMSprop optimizer and ReLU activation function.

### Model Accuracy

Training convergence was more effectively achieved with $L_{\mathrm{MSE}}$ (Fig. 7). This smoother convergence results from the linearity of $L_{\mathrm{MSE}}$, simplifying the optimization of model weights. In contrast, $L_{\mathrm{MSLE}}$, which introduces a logarithmic transformation to the loss calculation, presented challenges in convergence. The nonlinear scaling of $L_{\mathrm{MSLE}}$ complicates the optimization process, as it adjusts the error contributions in a nonuniform manner, making it more difficult to fine-tune the model's weights accurately.

Upon evaluating the models purely on testing data, the linear loss function exhibited a higher $R^2 = 0.967$ than the logarithmic loss function $R^2 = 0.86$ (Fig. 8). The values of $R^2_{\mathrm{adj}}$ were similar to regular $R^2$, indicating that adding more predictors did not substantially inflate the coefficient of determination. In contrast, SI differed substantially between the two approaches, reaching 0.926 for $L_{\mathrm{MSLE}}$ and only 0.503 for $L_{\mathrm{MSE}}$. This disparity reflects the tendency of the logarithmic loss function to more accurately capture low $N_{\mathrm{f}}$ samples, as seen in studies that reported lower fatigue life ranges, while the linear loss function better fits high $N_{\mathrm{f}}$ observations. Consequently, $L_{\mathrm{MSLE}}$ is advantageous when characterizing the distribution of fatigue lives at the lower end, even though $L_{\mathrm{MSE}}$ may yield a higher $R^2$ over the full data set.

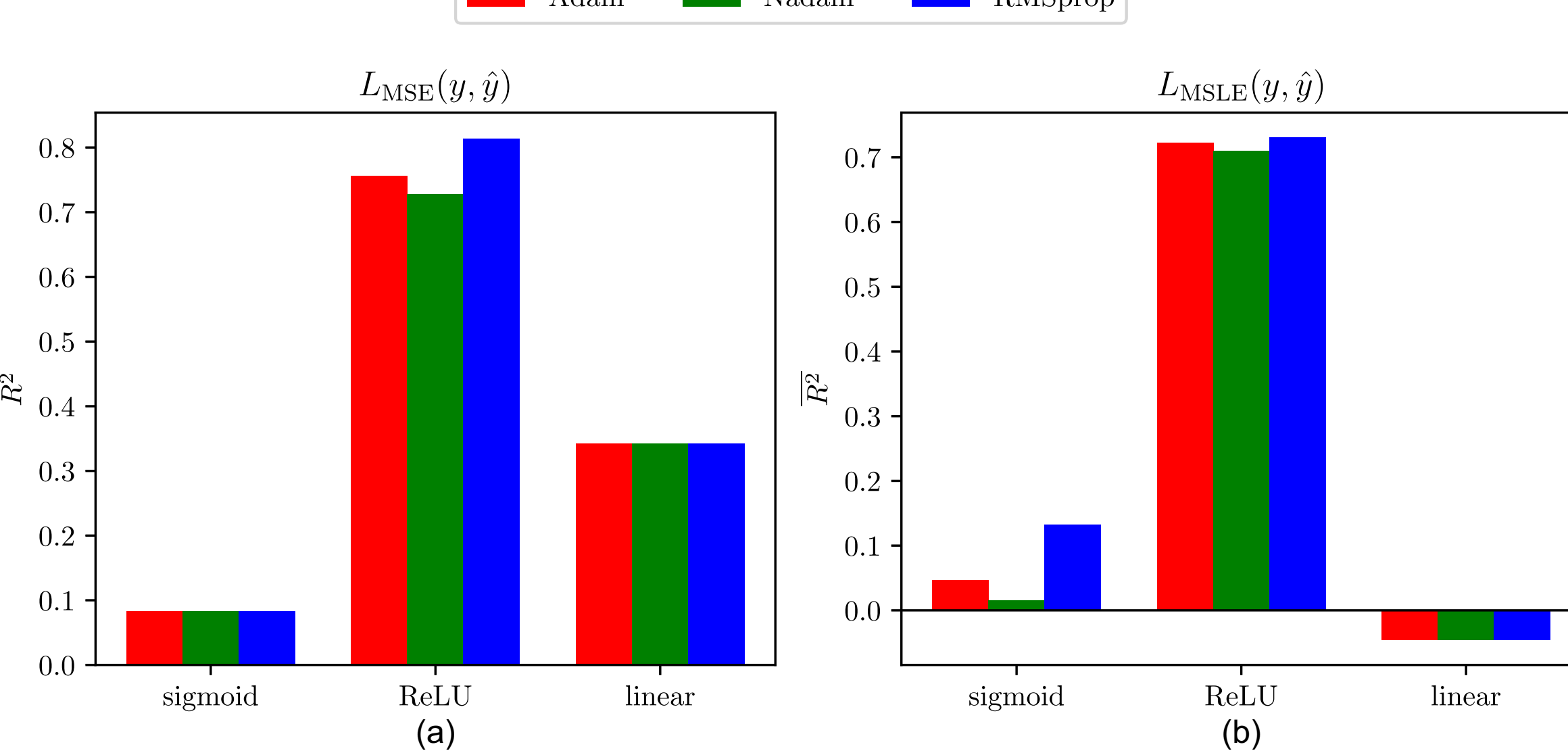

**Fig. 6.** Model performance, measured by the $\overline{R^2}$ score (mean for different folds), in relation to the activation functions and optimizers used, when employing (a) $L_{\mathrm{MSE}}$ and (b) $L_{\mathrm{MSLE}}$ as the loss functions; $n_{\mathrm{h}} = 2$ and $n_{\mathrm{neur,h}} = 200$.

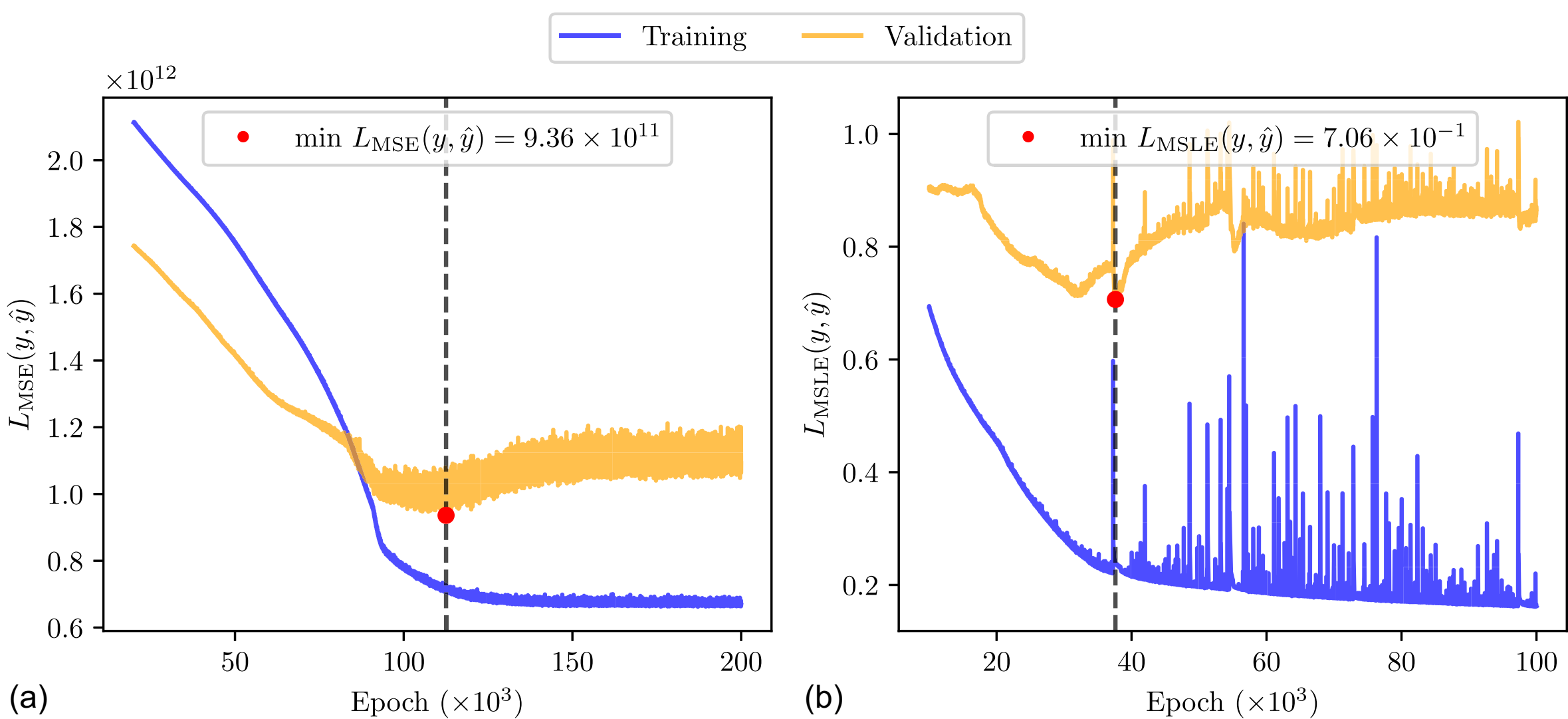

**Fig. 7.** Training convergence plots illustrating the progression of the training process over 200 000 epochs for: (a) $L_{\mathrm{MSE}}$ and (b) 100 000 epochs for $L_{\mathrm{MSLE}}$.

ANNs have been effectively applied to predict AC fatigue properties. A study by Isied and Souliman (2019) demonstrated an ANN model achieving $R^2 = 0.93$ for the entire data set, employing a simple architecture with notable predictive accuracy. However, this study used data exclusively sourced from the report by Witczak et al. (2013) and focused on predicting $\varepsilon$ based on $N_{\mathrm{f}}$ values. MATLAB was used to develop a three-layer feedforward backpropagation neural network with one hidden layer and a sigmoid activation function. The model utilized four input parameters: stiffness ratio at cycle number, initial stiffness, rest period, and number of cycles, to predict $\varepsilon$.

Xiao et al. (2009) explored ANN to predict fatigue properties, focusing on air voids and voids filled with asphalt, achieving high predictive accuracy ($R^2$ up to 0.97). Their data set, however, consisted of rubberized AC mixtures containing reclaimed asphalt pavement and was less complex, with lower variability compared to our data set. While their findings underscore the potential of ANN for predicting fatigue life, the reduced variability in their data may have contributed to the high accuracy, potentially leading to overfitting.

Additionally, Wang et al. (2023) compared traditional regression models for asphalt fatigue cycles, including, reaching the highest $R^2 = 0.78$. The relationship between initial stiffness and fatigue properties of asphalt mixes was investigated by Lundstrom et al. (2004), with a range of $R^2$ from 0.30 to 0.86 across different data sets.

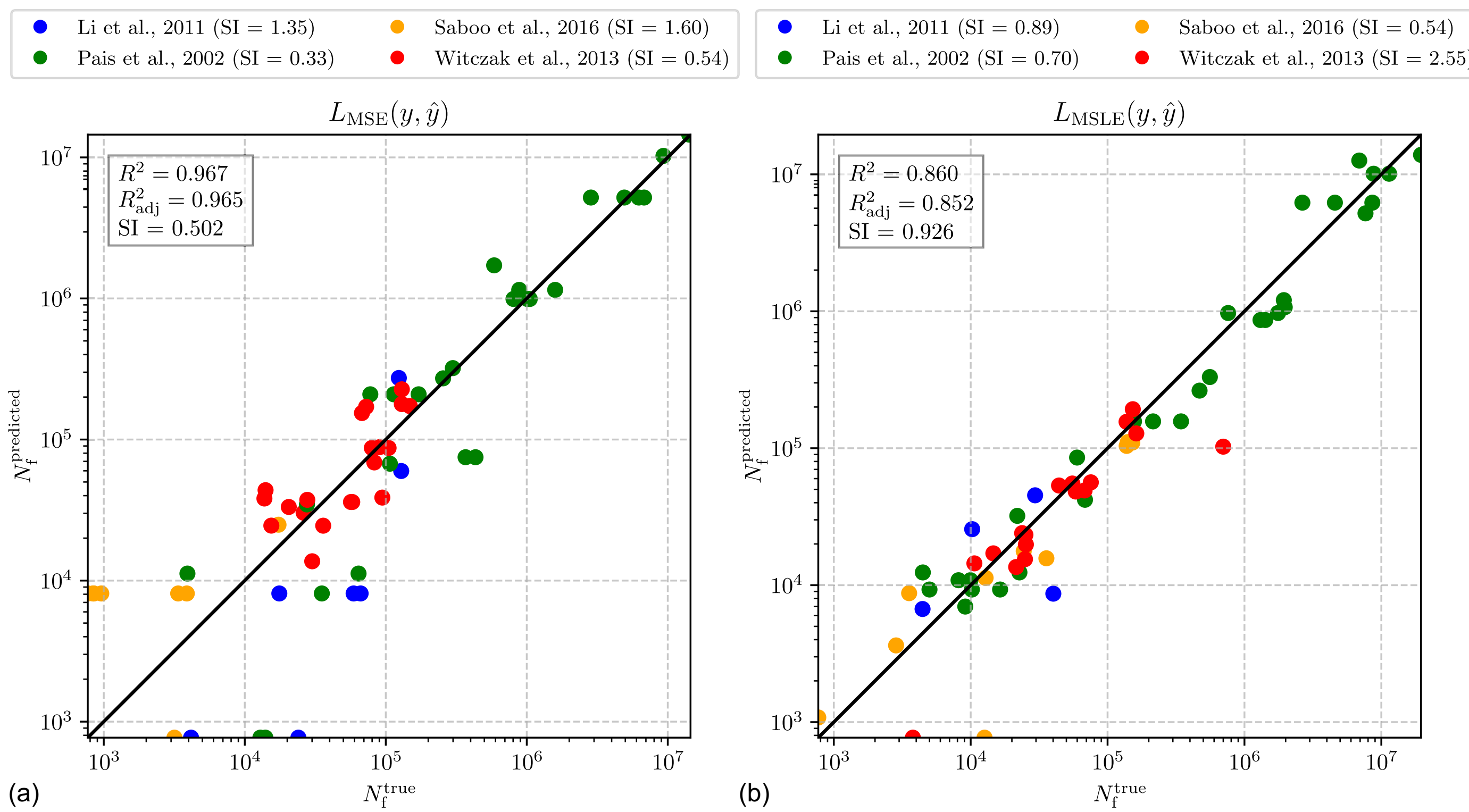

**Fig. 8.** Comparison of true versus predicted values for the testing data set when using: (a) $L_{\mathrm{MSE}}$ and (b) $L_{\mathrm{MSLE}}$; metrics shown include $R^2$, $R^2_{\mathrm{adj}}$, and SI, which is evaluated for whole testing data set and for individual studies separately in the legend.

Yan et al. (2018) reported on constructing an ANN for $N_f$ prediction. Their work explicitly details the key inputs, including asphalt content (4%, 5.5%, and 7%), air void levels (2.16%–6.22%), and strain levels (500–1,500 × $10^{-6}$), as well as testing conditions such as the 4PB fatigue test conducted at 5°C in strain-controlled mode. While their ANN model, featuring a three-layer feedforward structure with a single hidden layer, achieved high predictive accuracy $R^2 = 0.91$, the study was based on a limited data set of only 27 samples. This small sample size may limit the generalizability of their conclusions.

## Limitations

While a larger and more varied data set was utilized in this study, certain limitations must still be considered when interpreting the results. The data set is inherently imbalanced, with a substantial portion of the data contributed by a small number of studies. Although this may improve the model's ability to capture patterns characteristic of these dominant sources, it also raises concerns regarding potential overfitting and reduced generalizability, especially in the prediction of extreme $N_f$ values that are not well represented. In such cases, models trained on imbalanced data may exhibit a bias toward trends present in the more heavily sampled sources, leading to decreased accuracy for minority cases or rarer fatigue scenarios. Balancing the data set by reducing the number of samples from the larger studies was deliberately avoided to preserve the full variability in fatigue behavior. Instead, all available data were retained to ensure a comprehensive coverage of the response spectrum, with the understanding that this may introduce bias favoring the dominant sources and influence the model's performance in less-represented areas of the input space. Therefore, collecting new experimental data is essential to enhance model generalizability, and can be readily integrated using the public code in author's GitHub repository.

In addition to data set imbalance, several other methodological limitations should be noted. First, the overall data set remains relatively small, which limits the model's capacity to generalize across the full range of input and output values. Second, the use of a single neural network architecture to model the entire fatigue life spectrum may constrain predictive accuracy, particularly at the distribution tails. Future work could explore the development of range-specific models or ensemble techniques to better capture nonlinearities in different data regions. As the data set expands, additional input features, such as initial stiffness, binder type, or penetration grade, could be incorporated to improve model robustness. Furthermore, hyperparameters optimized for the current data set may need refinement as new data become available, highlighting the importance of continuous validation. These considerations point to the need for ongoing model adaptation and data set enrichment to strengthen the predictive performance and practical utility of the approach.

## Practical Outcomes

Further analysis involved generating partial dependence plots presented as 2D maps (Figs. 9 and 10) and as 2D line plots (Figs. 11 and 12) to explore the relationships between input parameters and predicted $N_f$ values for both models. These plots were constructed while adhering closely to the scatter distribution of the input data to avoid extrapolation, which could yield unreliable results. The input data points are marked with white triangular markers. Predictions were executed at two strain levels, 200 × $10^{-6}$ and 400 × $10^{-6}$, revealing that a lower air-void content (ideally below 3%) correlates with an extended fatigue life. In layers where fatigue is most critical, such as those using AC16 to AC22 mixes, void content typically ranges between 4% and 7%. In these cases, maintaining void content closer to 4% is generally preferable.

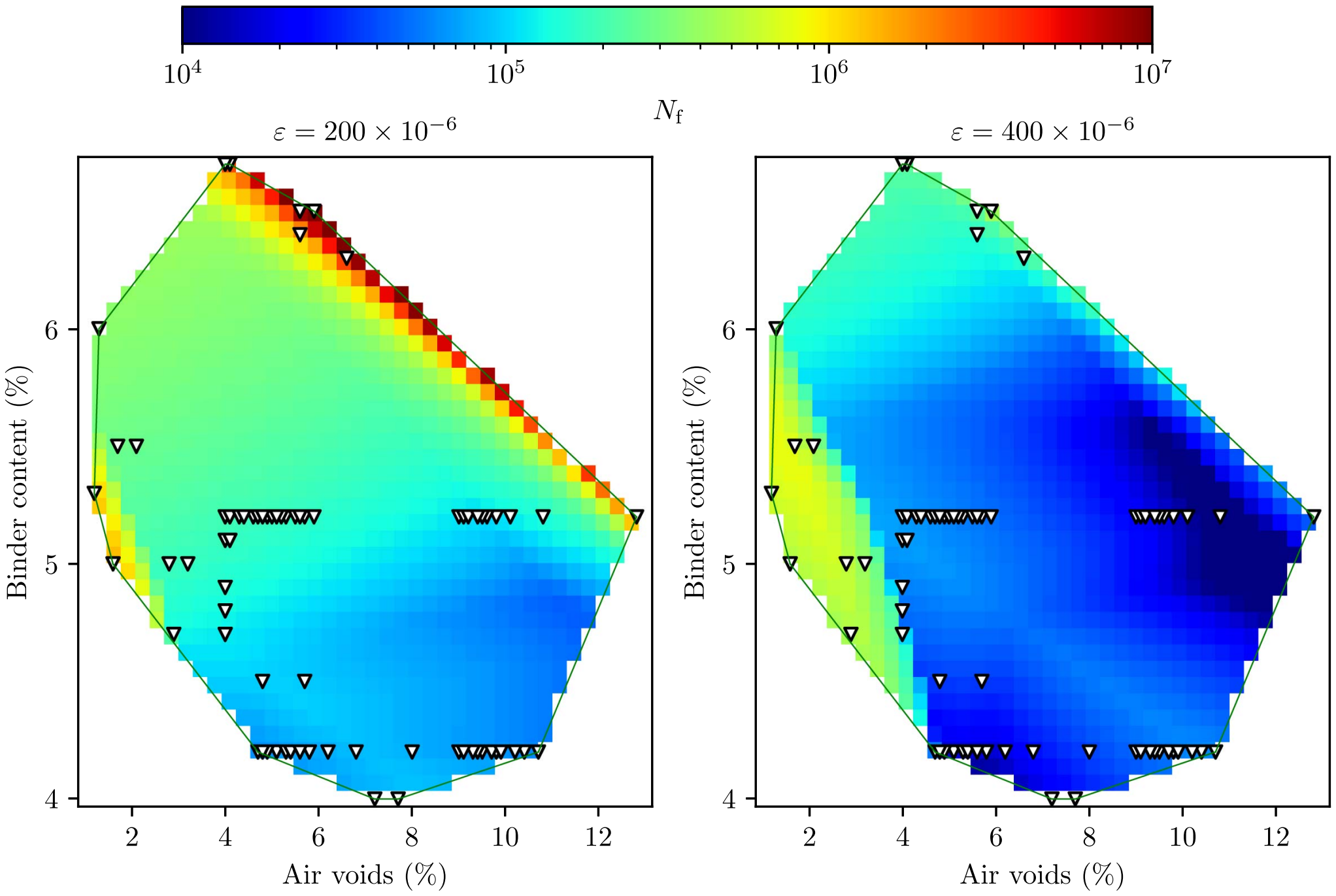

**Fig. 9.** Partial dependence plots generated using the calibrated model trained with $L_{\mathrm{MSE}}$, illustrating the relationship between input parameters and the predicted $N_{\mathrm{f}}$ at two strain levels (left and right plot); the data points within the convex hull of the measured data set are shown as white triangular markers.

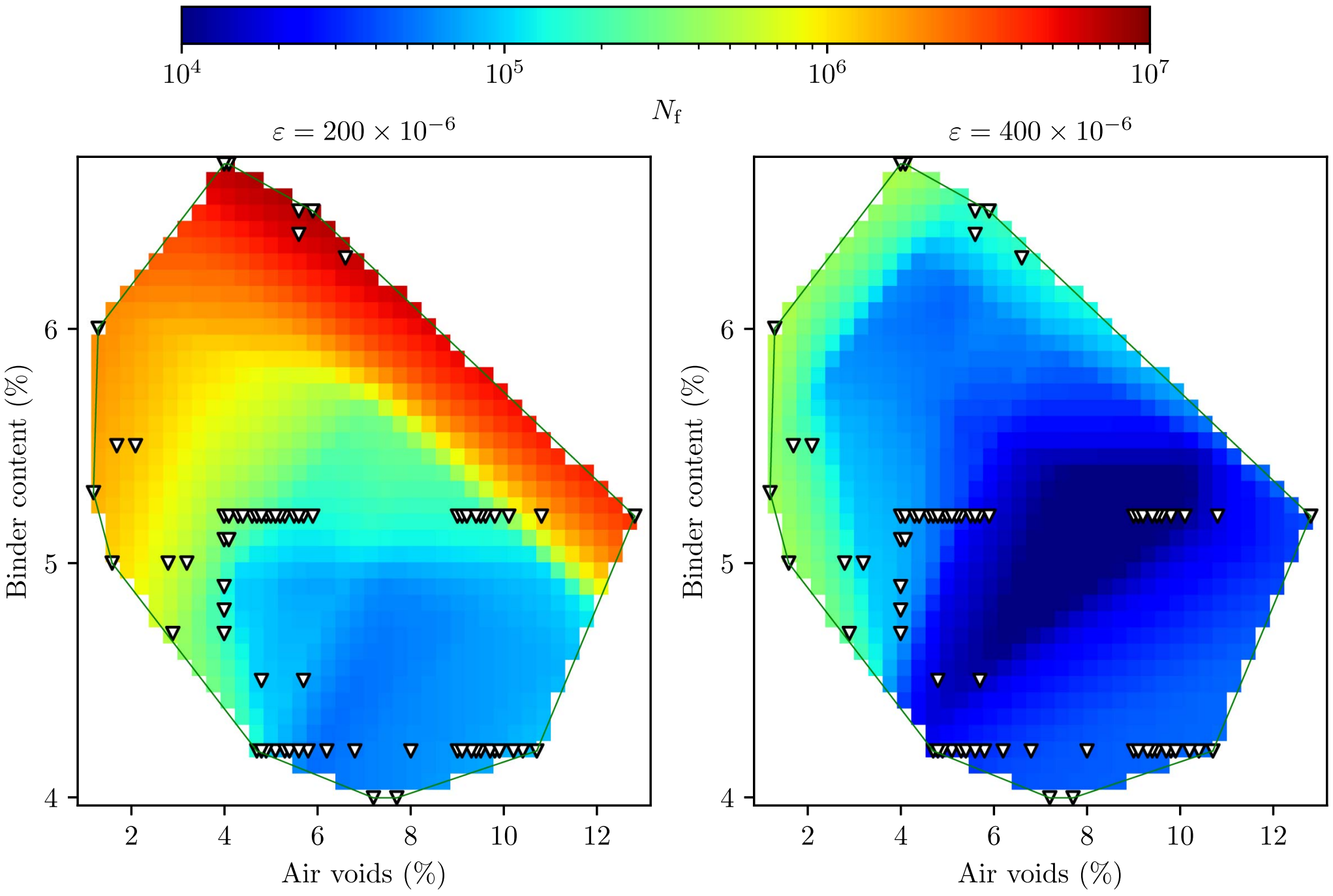

**Fig. 10.** Partial dependence plots generated using the calibrated model trained with $L_{\mathrm{MSLE}}$, illustrating the relationship between input parameters and the predicted $N_{\mathrm{f}}$ at two strain levels (left and right plot); the data points within the convex hull of the measured data set are shown as white triangular markers.

Similarly, an increase in binder content was found to enhance fatigue life, with $L_{\mathrm{MSLE}}$-based models indicating that high binder levels (over 5.5%) could mitigate the adverse effects of high air-void content.

Both models agree that a higher air-void content combined with a lower binder content results in AC with inferior fatigue life. These findings suggest that durable AC should contain at least 5.5% bituminous binder. However, this threshold may not be universally applicable, as AC22 mixtures used in subbase layers are typically designed with binder contents around 3.9%–4%, and values exceeding 6% may be excessive. A binder content above 5% is generally considered optimal for achieving a long-lasting mix. While the models indicate that higher air-void content could be advantageous when paired with a high binder content, this scenario falls outside the coverage of the input data, leading to extrapolation and requiring further experimental validation.

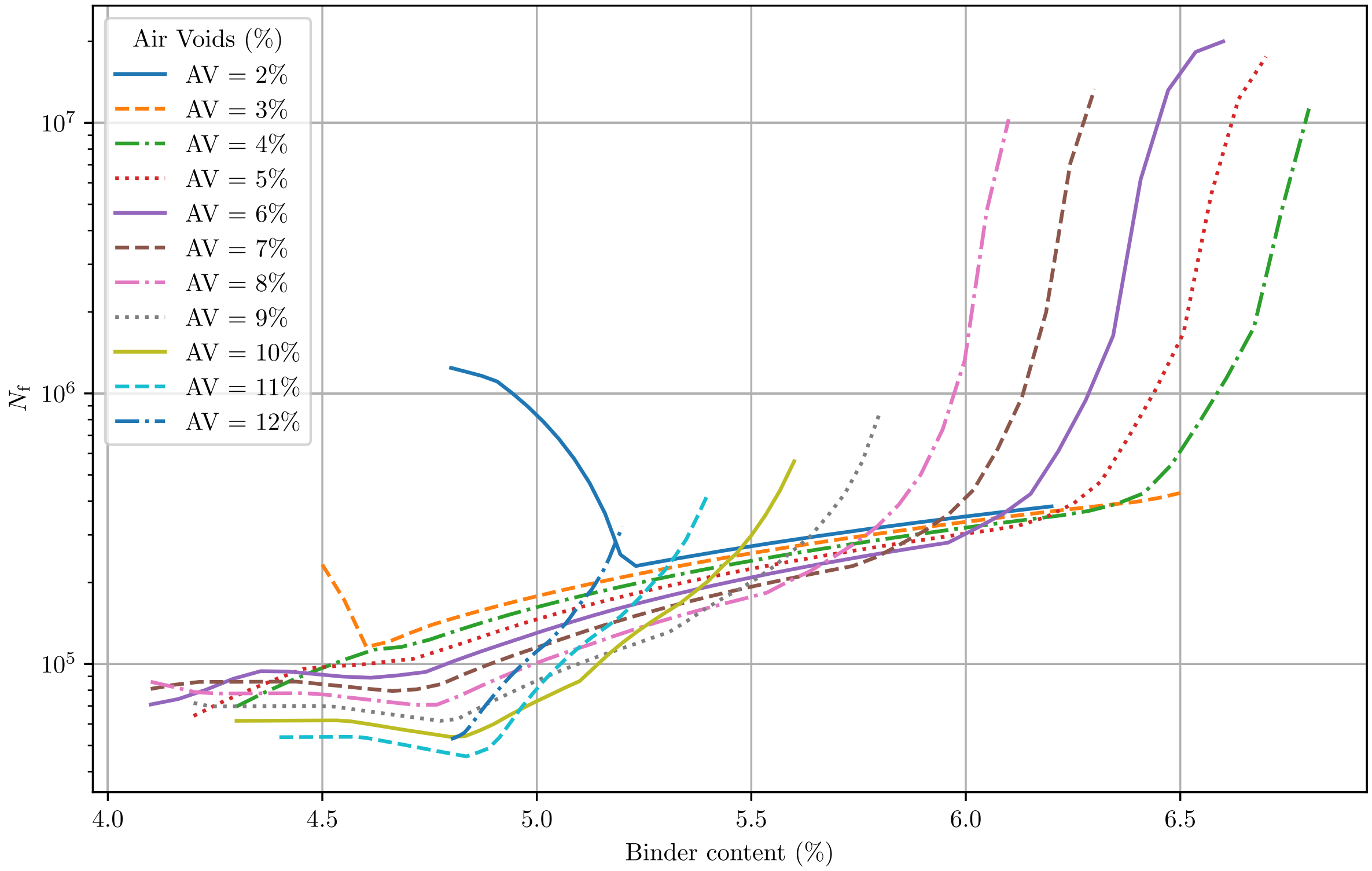

**Fig. 11.** 2D line plots representing sections of the partial dependence plots for $\varepsilon = 200 \times 10^{-6}$ within the convex hull of measured data, illustrating the relationship between two input parameters and the predicted $N_f$; the model was trained using $L_{\mathrm{MSE}}$.

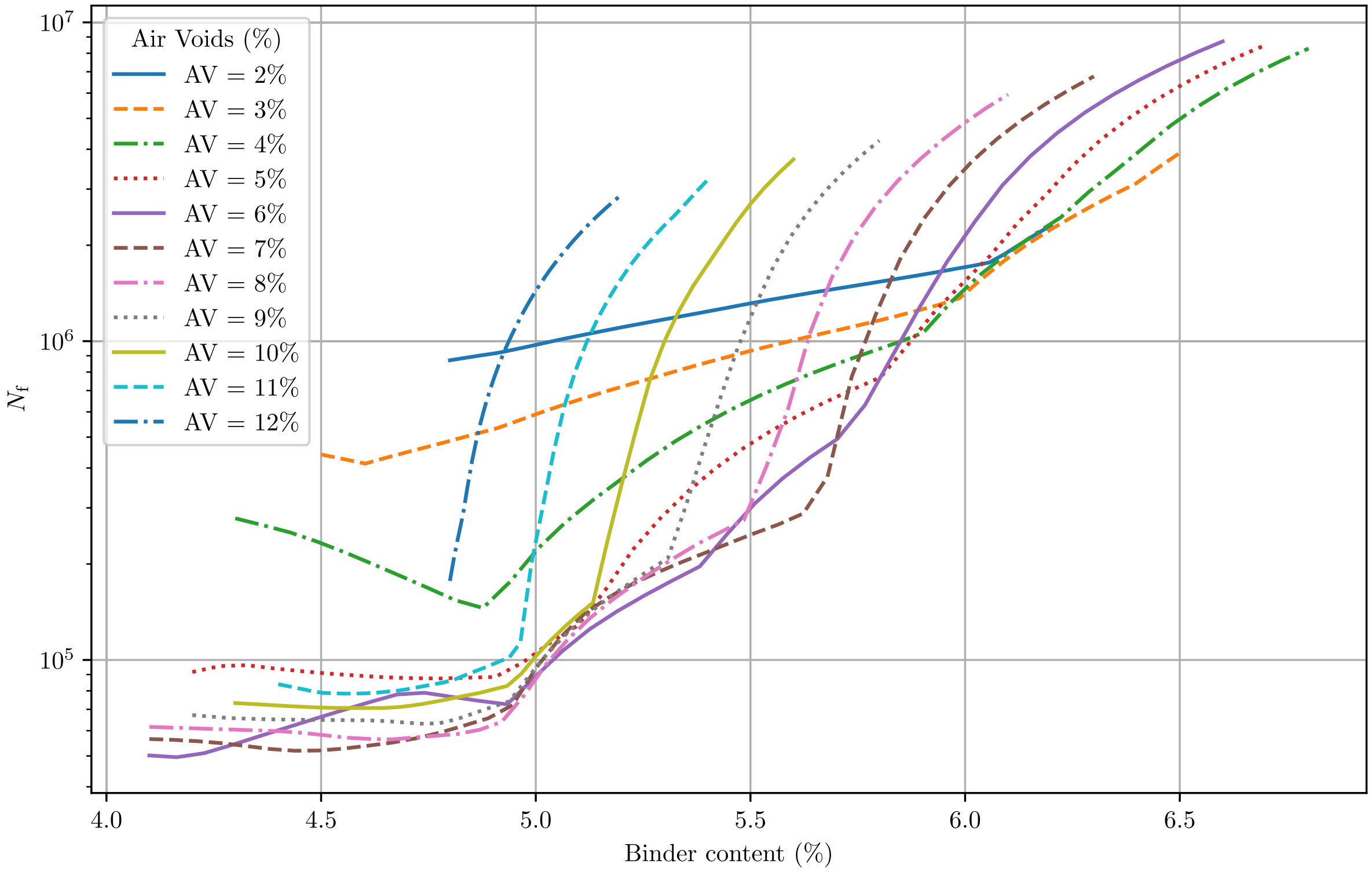

**Fig. 12.** 2D line plots representing sections of the partial dependence plots for $\varepsilon = 200 \times 10^{-6}$ within the convex hull of measured data, illustrating the relationship between two input parameters and the predicted $N_f$; the model was trained using $L_{\mathrm{MSLE}}$.

These findings align with the trends reported by Zeiada et al. (2013), who observed higher $N_f$ values with an increase in binder content from 4.2% to 5.2%. Additionally, they noted that increasing air-void content from 4.5% to 9.5% resulted in relatively similar average fatigue life but with higher variability. While the data from Zeiada et al. (2013) were not used as inputs for our models due to differences in testing methods, the qualitative trends they identified provide valuable context for validating the relationships revealed by our predictions.This phenomenon was also scrutinized by Ma et al. (2016), who proved in their numerical model that AC specimens having the same air-void content can exhibit different fatigue life due to the nonuniform distribution of air voids. Our findings are aligned with the results of Yan et al. (2018) in that $N_f$ increases with binder content, while the impact of air voids is less pronounced.

High air-void content weakens cohesion between aggregate particles, accelerates fatigue failure, and increases susceptibility to moisture and oxygen infiltration, leading to binder aging and structural degradation. However, excessively low voids raise the risk of rutting, as asphalt mortar requires sufficient space for stress redistribution under repeated loading.

Binder content is crucial for fatigue resistance, enhancing adhesion, cohesion, and moisture protection while slowing aging effects. Its viscoelastic properties improve stress dissipation, reducing crack propagation. Conversely, insufficient binder weakens aggregate bonding and increases fatigue susceptibility, particularly in base layers subjected to cyclic bending. Rich bottom layers with slightly increased binder content (0.5%–1% by mass) are often used in perpetual pavements to enhance durability.

Beyond air voids and binder content, other factors influence fatigue performance. The initial stiffness of the mixture, especially when using hard or polymer-modified binders, significantly affects fatigue resistance. High-modulus asphalt concrete, commonly applied in heavily loaded pavements, mitigates fatigue cracking by optimizing layer thickness (typically 10–12 cm). Incorporating binder stiffness into predictive models could improve fatigue predictions, but requires additional data, such as penetration and binder fatigue characteristics. Additionally, bitumen mortar properties, which include binder-filler interactions, may play a key role.

At its current stage of development, the model serves as a practical tool for the preliminary estimation of expected $N_f$ values across different strain levels. This enables engineers to perform early-stage evaluations and adjust design parameters to achieve desired fatigue performance without immediate reliance on extensive experimental testing. Although caution must be exercised when interpreting predictions at the extremes of the data set, the model provides valuable guidance in general fatigue regimes, especially for mixes resembling the dominant data sources used in training. Looking ahead, this modeling approach has the potential to reduce the burden of traditional testing protocols, which require numerous specimens and are highly time- and resource-intensive. With further refinement and expanded data sets, the model could support the development of more efficient testing strategies or even serve as a decision-support tool in pavement design frameworks.

## Conclusion

The study employed ANNs to predict the fatigue life of AC, focusing on the impact of strain level, binder content, and air voids. Utilizing a substantial data set, we tailored the ANN models to accommodate a wide range of fatigue life data, represented typically on a logarithmic scale. Our strategic use of both $L_{MSE}$ and $L_{MSLE}$ as loss functions allowed for fine-tuning the model's sensitivity to the scale of the data, addressing the logarithmic distribution of fatigue life.

Through rigorous hyperparameter optimization, we identified that a model architecture with two hidden layers containing 200 neurons provided the best performance. The ReLU activation function proved optimal, and the RMSprop optimizer emerged as the most stable.

Our analysis showed that $L_{MSE}$ reached a higher $R^2$ value on testing data, indicating stronger overall alignment with high-$N_f$ observations. However, $L_{MSLE}$ produced a significantly lower SI in the low-$N_f$ region, thus better preserving the relative order among smaller fatigue life values. These findings highlight that $L_{MSLE}$ is more effective for heterogeneous data sets containing substantial variation at lower fatigue life ranges, while $L_{MSE}$ may be preferable when emphasis is placed on performance in the higher range of $N_f$.

The detailed analysis of the developed models revealed critical insights into the relationships between AC's microstructural parameters and fatigue life. Lower air-void and higher binder content were associated with extended fatigue life. On the other hand, high air-void and low binder content markedly reduced fatigue life, underscoring the critical balance needed for durable AC. Such findings highlight the importance of optimizing binder and air-void content in AC mix design to improve pavement durability and reduce maintenance costs. Naturally, this must be balanced with other aspects of asphalt mix performance, such as resistance to permanent deformation (rutting) and susceptibility to thermal cracking.

### Data Availability Statement

Data, models, and code generated and used during the study are available in a repository online (https://github.com/VaclavNezerka/AsphaltFatigueANN) in accordance with funder data retention policies.

### Acknowledgments

During the preparation of this work, the authors used AI language model ChatGPT 4 from OpenAI L.L.C. (CA 94110 San Francisco, USA) to improve readability and language of the manuscript. After using these tools, the authors reviewed and edited the content as needed, and take full responsibility for the content of the publication. This work was supported by the Czech Science Foundation (No. 22-04047K) and the Czech Technical University in Prague, Grant Agreement No. SGS25/005/OHK1/1T/11. V. Nežerka's contribution was funded by the European Union under the project ROBOPROX (No. CZ.02.01.01/00/22_008/0004590).

### Author Contributions

Jakub Houlík: Data curation; Investigation; Software; Visualization; Writing – original draft. Jan Valentin: Conceptualization; Funding acquisition; Investigation; Project administration; Supervision; Writing – review and editing. Václav Nežerka: Conceptualization; Methodology; Software; Supervision; Visualization; Writing – review & editing.